\documentclass[twocolumn,secnumarabic,amssymb, nobibnotes, aps, pra, superscriptaddress]{revtex4-2}
\usepackage{CJK}
\usepackage{graphicx}
\usepackage{dcolumn}
\usepackage{bm}
\usepackage{physics}
\usepackage[utf8]{inputenc}
\usepackage{color}
\usepackage[T1]{fontenc}
\usepackage{booktabs, array, mathptmx, float, tabularx, booktabs, lipsum, amsmath,multirow}
\usepackage{siunitx, xcolor}
\graphicspath{{figs/}{figsgaoerb/}} 
\usepackage[colorlinks,linkcolor=blue,anchorcolor=blue,citecolor=blue,urlcolor=blue]{hyperref}

\begin{document}


\title{Estimation of the Laser Frequency Nosie Spectrum  by Continuous Dynamical Decoupling}


\author{Manchao Zhang}
\thanks{These authors contributed equally to this paper.}
\affiliation{ Department of Physics, College of Liberal Arts and Sciences, National University of Defense
Technology, Changsha 410073, Hunan, China}
\affiliation{  Interdisciplinary Center for Quantum Information, National University of Defense Technology, Changsha 410073, Hunan, China}

\author{ Yi Xie}
\thanks{These authors contributed equally to this paper.}
\affiliation{ Department of Physics, College of Liberal Arts and Sciences, National University of Defense
Technology, Changsha 410073, Hunan, China}
\affiliation{  Interdisciplinary Center for Quantum Information, National University of Defense Technology, Changsha 410073, Hunan, China}

\author{Jie Zhang}
\affiliation{ Department of Physics, College of Liberal Arts and Sciences, National University of Defense
Technology, Changsha 410073, Hunan, China}
\affiliation{  Interdisciplinary Center for Quantum Information, National University of Defense Technology, Changsha 410073, Hunan, China}

\author{Weichen Wang}
\affiliation{ Department of Physics, College of Liberal Arts and Sciences, National University of Defense
Technology, Changsha 410073, Hunan, China}
\affiliation{  Interdisciplinary Center for Quantum Information, National University of Defense Technology, Changsha 410073, Hunan, China}

\author{Chunwang Wu}
\affiliation{ Department of Physics, College of Liberal Arts and Sciences, National University of Defense
Technology, Changsha 410073, Hunan, China}
\affiliation{  Interdisciplinary Center for Quantum Information, National University of Defense Technology, Changsha 410073, Hunan, China}

\author{Ting Chen}
\affiliation{ Department of Physics, College of Liberal Arts and Sciences, National University of Defense
Technology, Changsha 410073, Hunan, China}
\affiliation{  Interdisciplinary Center for Quantum Information, National University of Defense Technology, Changsha 410073, Hunan, China}

\author{Wei Wu}
\email{weiwu@nudt.edu.cn}
\affiliation{ Department of Physics, College of Liberal Arts and Sciences, National University of Defense
Technology, Changsha 410073, Hunan, China}
\affiliation{  Interdisciplinary Center for Quantum Information, National University of Defense Technology, Changsha 410073, Hunan, China}

\author{Pingxing Chen}
\email{pxchen@nudt.edu.cn}
\affiliation{ Department of Physics, College of Liberal Arts and Sciences, National University of Defense
Technology, Changsha 410073, Hunan, China}
\affiliation{  Interdisciplinary Center for Quantum Information, National University of Defense Technology, Changsha 410073, Hunan, China}


\date{\today}

\begin{abstract}
Decoherence induced by the laser frequency noise is one of the most important obstacles in the quantum information processing. In order to suppress this decoherence, the noise power spectral density needs to be accurately characterized. In particular, the noise spectrum measurement based on the coherence characteristics of qubits would be a meaningful and still challenging method. Here, we theoretically analyze and experimentally obtain the spectrum of laser frequency noise based on the continuous dynamical decoupling technique. We first estimate the mixture-noise (including laser and magnetic noises) spectrum  up to $(2\pi)$530 kHz  by monitoring the transverse relaxation from an initial state $+X$, followed by a gradient descent data process protocol.  Then the contribution from the laser noise is extracted by enconding the qubits on different Zeeman sublevels. We also investigate two sufficiently strong noise components by making an analogy between these noises and driving lasers whose linewidth assumed to be negligible. This method is verified experimentally and finally helps to characterize the noise.

\end{abstract}


\maketitle

\section{Introduction}
\label{sectionI}
The problem of a quantum system interacting with a noisy environment is of great significance in the field of quantum computing\cite{arute2019quantum}. In general, there are some strategies to fight this decoherence and improve the fidelity of quantum operations. One method is based on the error-correction protocols by encoding multiple physical qubits into a logical qubit, which  of course requires tremendous amount of qubit resources\cite{ofek2016extending}. And the second method is  reducing the amplitude of environmental noise. This can be realized by applying a reverse compensation  signal targeting the interested noise\cite{Merkel2019Magnetic}. Or in contrast, reduce the system's internal sensitivity to noise through the application of coherent control-pulse methods\cite{Gullion1990New,Ryan2010Robust,ajoy2011optimal,Souza2011Robust,LidarDemonstration,wang2017single-qubit} or encoding the information to pairs of dressed states which are formed by concatenated continuous driving fields\cite{Bermudez2011Robust,Cai2012Robust,Farfurnik2017Experimental,Xiangkun2012Coherence,Timoney2011Quantum,D2014Protecting,Stark2017Narrow}. However, the latter two approachs depend on the prior knowledge of the noise  power spectral density(PSD) \cite{christensen2019anomalous,chan2018assessment}.

Laser frequency noise or phase noise describes how the frequency of a laser  field deviates from an ideal value. This quantity, which is defined to evaluate the short-term stability of a laser, has attracted widespread attention as a fundamental topic about lasers. Generally, the noise features of a laser  can be revealed through the following two schemes. One is the optical heterodyne method which contains two different categories. The first one is comparing the laser with itself through the delayed self-heterodyne scheme\cite{2013Phase}. For an ultra-narrow-linewidth laser, at least hundreds of kilometers of optical fiber are required to realize a delay time longer than the laser coherent time.  The other one is comparing the frequency of a target laser with reference lasers by the light-beam heterodyne method, which is also called as beat-note  method in some articles\cite{walsh2012characterization}. If the target laser is beat with only one reference laser, we will get an electrical signal that contains noise information of both lasers, but fail to separate their contributions. Instead, if two reference lasers are introduced in the experiment, two eletrical signals can be detected. These signals are first mixed down to a lower frequency and then analyzed by a digital cross-correlation method to characterize the frequency noise PSD of the target laser\cite{xie2017phase}. In other words, to obtain the noise PSD, we have to set up at least other two similiar lasers. This is uneconomic and has limited applications. The other method is refering the laser frequency to the atomic energy levels. The Rabi spectrum protocol, for example, can be used to the PSD  characterization by using a weak and long excitation pulse. But the result can only be used as a reference as the accuracy is insufficient.

Another simple and effective  technique for the noise PSD estimation is called pulsed dynamical decoupling (PDD)\cite{yuge2011measurement,bylander2011noise,norris2018optimally,cappellaro2006principles,sung2019non-gaussian}. PDD consists of the applying of $\pi$ pulses, $X_\pi$ or $Y_\pi$, to the system with  designed intervals, which counteracts the coherence decay due to the noise, such as the Carl-Purcell-Meiboom-Gill (CPMG)\cite{Almog2011Direct} and Uhrig dynamical decoupling (UDD)\cite{Mukhtar2010Protecting} sequences. In general, beside the laser frequency noise, laser power fluctuations and magnetic field fluctuations are also very important noise sources in quantum information processing (QIP), especially for many solid-state qubits. If the laser power fluctuations are suppressed in the experimental setup, the rest two single-axis noises will be dominant, and affect the qubit evolution in a similiar way\cite{zhang2020discrimination}. Based on different quantum systems, the PDD technique has been utilized to  extract the contributions of single magnetic noise\cite{alvarez2011measuring,wang2017single-qubit} or laser frequency noise\cite{Bishof2013Optical}. However, for the practical applications, there also exist some drawbacks of PDD. On one hand, the finite $\pi$-pulse length can not be ignored in the experiments, and this will complicate the derivation of filter function which is the key tool for extracting noise spectrum.\cite{Wang2012Effect}. On the other hand, a large number of imperfect, finite-width pulses provoke the accumulation of errors (such as the pulse width error) and degrade the PDD performance\cite{He2018Effects}. Although these errors can be compensated or eliminated by additional sequence design and operations, the complexity of PDD increases\cite{Bishof2013Optical}. This makes the protocol inefficient when the fidelity of a single $\pi$ pulse operation is not qualified.

An alternative approach is continuous dynamical decoupling (CDD) technique where a continuous driving field is used instead of the $\pi$ pulses in PDD\cite{stark2017narrow-bandwidth,yan2013rotating-frame}. Compared to PDD, the concise, unimodal filter function as well as simple, high fidelity operations ensure better performance of CDD. In this paper, we mainly introduce the first applying of the CDD approach to the spectral estimation of laser frequency noise  in a trapped ion system. We also propose  a "laser-like noise" method  (see section \ref{sectionIII}) for the characterization of two strong components as well as the frequency region around them  of laser noise in our system. Note that compared to the conventional optical heterodyne approach, these methods are actually handy and economical for characterizing the interested manipulation lasers. And the laser noise spectrum  obtained here can be used for: (i) prolonging the coherence time and improving manipulation fidelity by choosing appropriate experimental parameters; (ii) improving the laser performance  by analyzing and then controlling the noise sources.

The remainder of this paper is organized as follows. Section \ref{sectionII} reviews the fundamental of CDD protocol and provides the control sequences in the experiments. In section \ref{sectionIII}, we propose and experimentally validate a new noise characterization scheme adapted to the strong noise situation. Then the main result of this paper is given in section \ref{sectionIIII} where the PSD of laser frequency noise is obtained. In section \ref{sectionIIIII}, we compare the noise spectra based on our scheme with beat-note results. And finally, we conclude in section \ref{sectionIIIIII}.

\section{Theoretical model and Control Setting}
\label{sectionII}
\begin{figure}[h]
\centering
\includegraphics[width=0.35\linewidth]{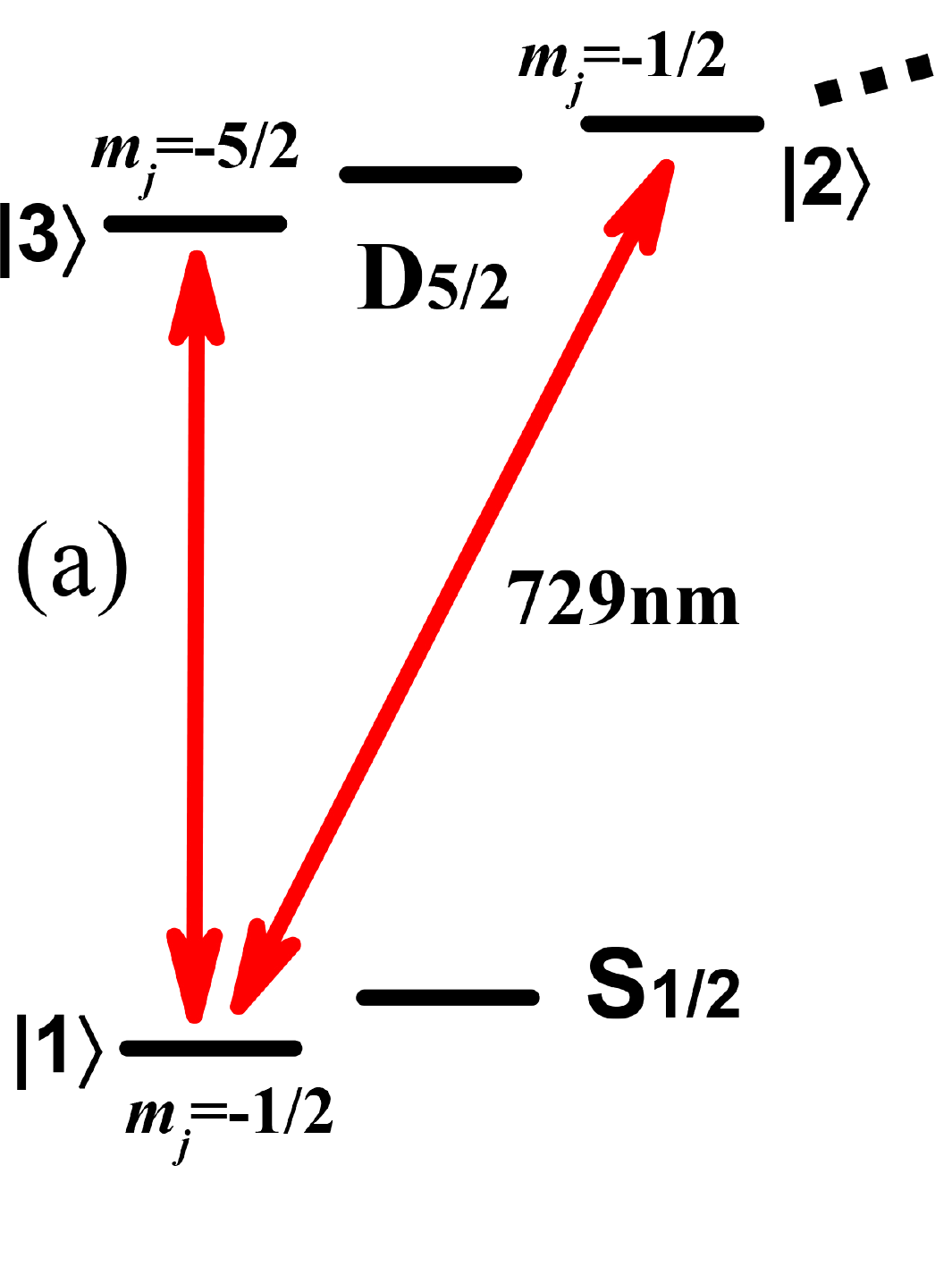}%
\includegraphics[width=0.65\linewidth]{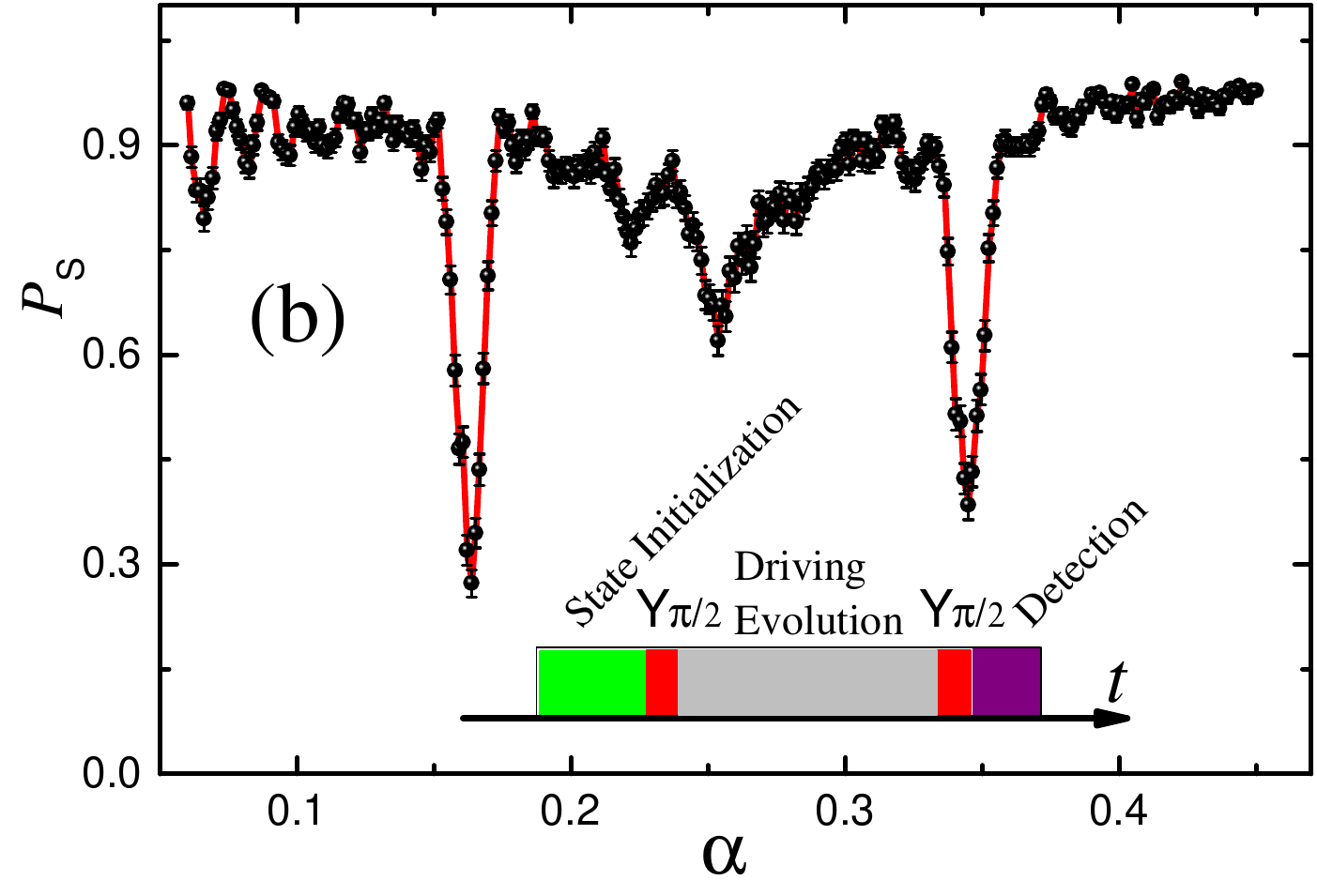}%
\caption{\label{Fig 1}{Level structure and control sequence. (a) Scheme of two level systems driven by 729 nm lasr. (b) Survival probability $P_s$ as a function of 729 nm-laser strength $\alpha$ for $t = 200 \rm{\mu s}$ and transition $\ket{1} \leftrightarrow \ket{3}$. By checking $\alpha$ with $\Omega$, we find $P_s$ has two local minimum values smaller than 0.5 at $\Omega \approx (2\pi)82$ kHz and $(2\pi)164$ kHz. They can be viewed as two sufficiently strong noise components which do not follow the rules of Eq. (\ref{Ps}) and should be modelled in another way. It should be mentioned that the linewidth around these components simply represents a power broadening as well as Fourier limit instead of the linewidth of the electrical signal driving etalon. Each point represents 200 experiments, and the error bars denote the projective measurement errors.}}
\end{figure}

For the present experiment we use a single ${}^{40}Ca^+$ ion confined in a standard blade trap. A pair of Zeeman sublevels in the ${}^2S_{1/2}$ and ${}^2D_{5/2}$ manifolds are  used as a qubit, as shown in FIG. \ref{Fig 1}(a). We notice that the magnetic fluctuation induced decoherence is distinguishable for different qubit definition, so we employ transitions $\ket{1}=\ket{{}^2S_{1/2}, m_j=-1/2} \leftrightarrow \ket{2}=\ket{{}^2D_{5/2}, m_j=-1/2}$ and   $\ket{1} \leftrightarrow \ket{3}=\ket{{}^2D_{5/2}, m_j=-5/2}$ as magnetic noise and thus laser noise probes, on which the same pulse sequences (as the inset in FIG.  \ref{Fig 1}(b)) are performed. In the experiment, we first initialize the qubit  to state $\ket{1}$ by the  Doppler cooling, sideband cooling as well as optical pumping methods. Then a $\pi/2$ $\sigma_y$ $(Y_{\pi/2})$  pulse is applied to rotate the qubit by $90^\circ$ to $+\sigma_x$ axis. During the driving evolution process, the 729 nm laser beam along $\sigma_x$ is turned on for time $t$, followed by an another $\pi/2$ $\sigma_y$ pulse. The final state is discriminated using a floresence detection scheme.

For the driving evolution steps, we consider a two-level system (TLS) subjected to the  noises  from  manipulation laser and magnetic field fluctuations. For resonant driving along the $\sigma_x$ axis with Rabi frequency $\Omega$, the Hamiltonian discribing the qubit system in the rorating frame can be written as ($\hbar= 1$)
\begin{equation}
\label{Hamiltonian1}
H = f(t)[\cos{( \Omega t)}\sigma_z + \sin{(\Omega t)}\sigma_y]/2.
\end{equation}
Here, $f(t)=f_B(t)-f_L(t)$ (unit: rad) is the mixture noise in time domain, and $f_B(t)(f_L(t))$ represents the part from magnetic field (laser frequency) fluctuations. Generally, $f(t)$ can be treated as a stationary, Gaussian-distributed function with zero mean, i.e., $\left\langle f(t)\right\rangle = 0$. The statistics of this stochastic process $f(t)$ is  fully determined by its autocorrelation function $C(t_1-t_2) = \left\langle f(t_1)f(t_2)\right\rangle$, or equivalently by its power spectral density  defined by the Fourier transform of $C(t)$: $S(\omega) = \int_{-\infty}^\infty C(t)e^{-i\omega t}dt$  (Wiener-Khinchin theorem).

In the weak noise limit, the connection between the driving evolution time $t$ and the survival probability $P_s$ of qubit on $+X$ state can be derived based on the stochastic Liouville theory and superoperator formalism\cite{willick2018efficient}. When only up to the second order cumulant is considered:
\begin{equation}
\label{Ps}
P_s(t) = \frac12+ \frac12 \exp(- \int_0^\infty d\omega S(\omega) F(\omega, \Omega, t)),
\end{equation}
where $F(\omega, \Omega, t) = t^2[\rm{sinc^2}$$(( \omega + \Omega) t/2) + \rm{sinc^2}$$((\omega - \Omega) t/2)]/4\pi$ is the filter function characterized by $t$, $\Omega$ and peaked at $\omega = \pm \Omega$ with full width at half maxumum (FWHM) $2\pi/t$ rad. This means the performance of the sequence is sensitive to the characteristics of the noise, and we can get some sketchy but impressive information about the noise strength at Fourier frequency $\omega =\Omega$ by simply scaning $\Omega$.

With $t = 200 $ $\rm{\mu s}$, we obtain the survival probability $P_s$ as a function of $\alpha$, which refers to the 729 nm-laser  amplitude modulation  used in arbitrary wave-form generator (AWG). As shown in FIG.  \ref{Fig 1}(b), there are two dominant noise components approximately at $ (2\pi)82$ kHz ($\alpha=0.1635$) and $(2\pi)164$ kHz ($\alpha=0.3450$) that are too strong to be described  by Eq. (\ref{Ps}). According to Eq. (\ref{Ps}), we will see an exponential decay from 1 to 0.5  with a time-dependent rate $\gamma(t) = \frac1t\int_0^\infty d\omega S(\omega) F(\omega, \Omega, t)$. Generally, this is valid only if the PSD of noise varies gently as a function of  $\omega$. However, FIG. \ref{Fig 1}(b) suggests that when $\Omega \approx (2\pi)82$ kHz or $(2\pi)164$ kHz, $P_s$ can be at least down to 0.3, smaller than the limit value 0.5 of Eq. (\ref{Ps}). This means that we have to research on these two components of noise by another method.

\section{Dominant noise characterization}
\label{sectionIII}

\begin{figure}[h]
\centering
\includegraphics[width=0.488\linewidth]{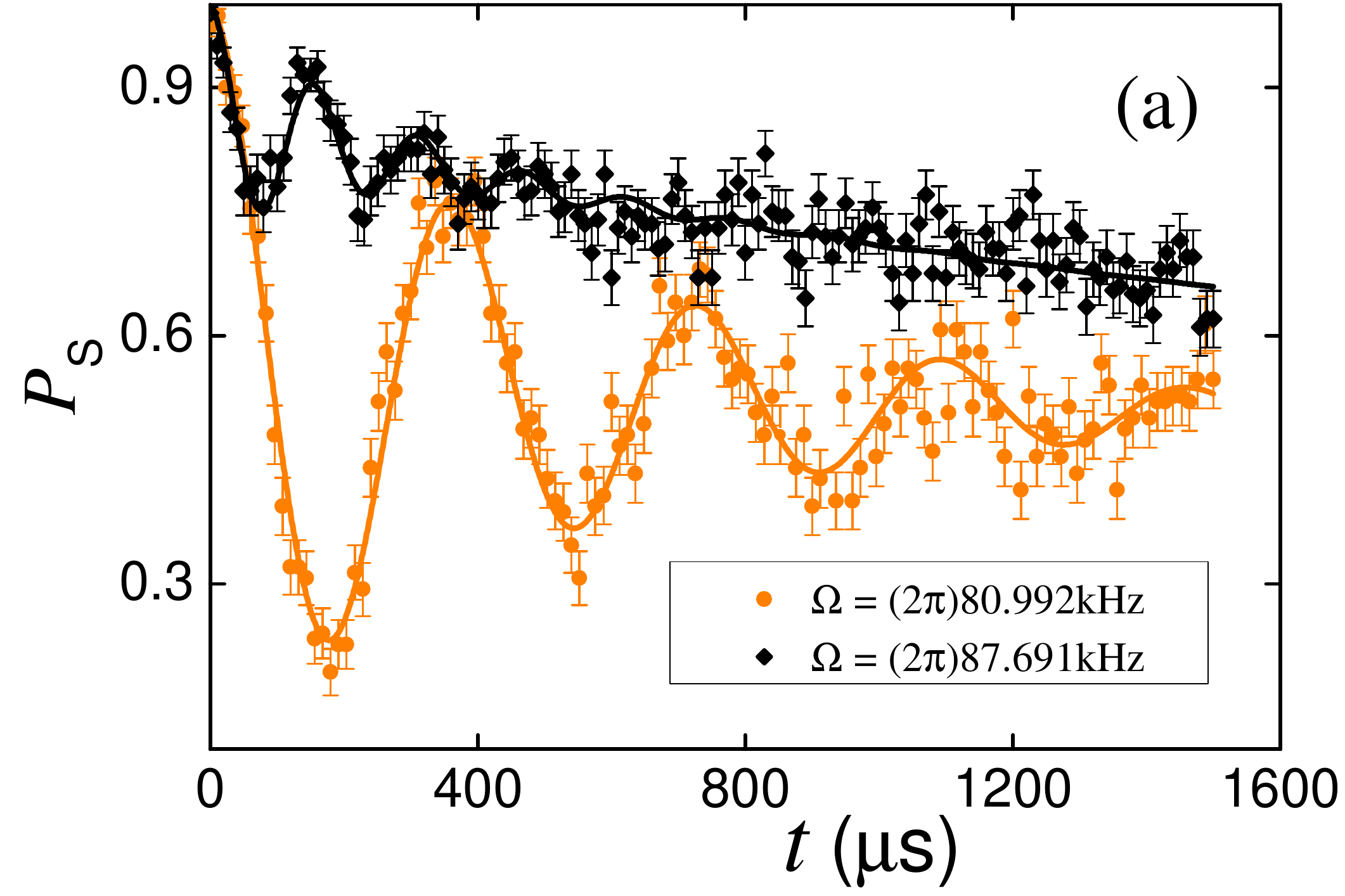}%
\includegraphics[width=0.512\linewidth]{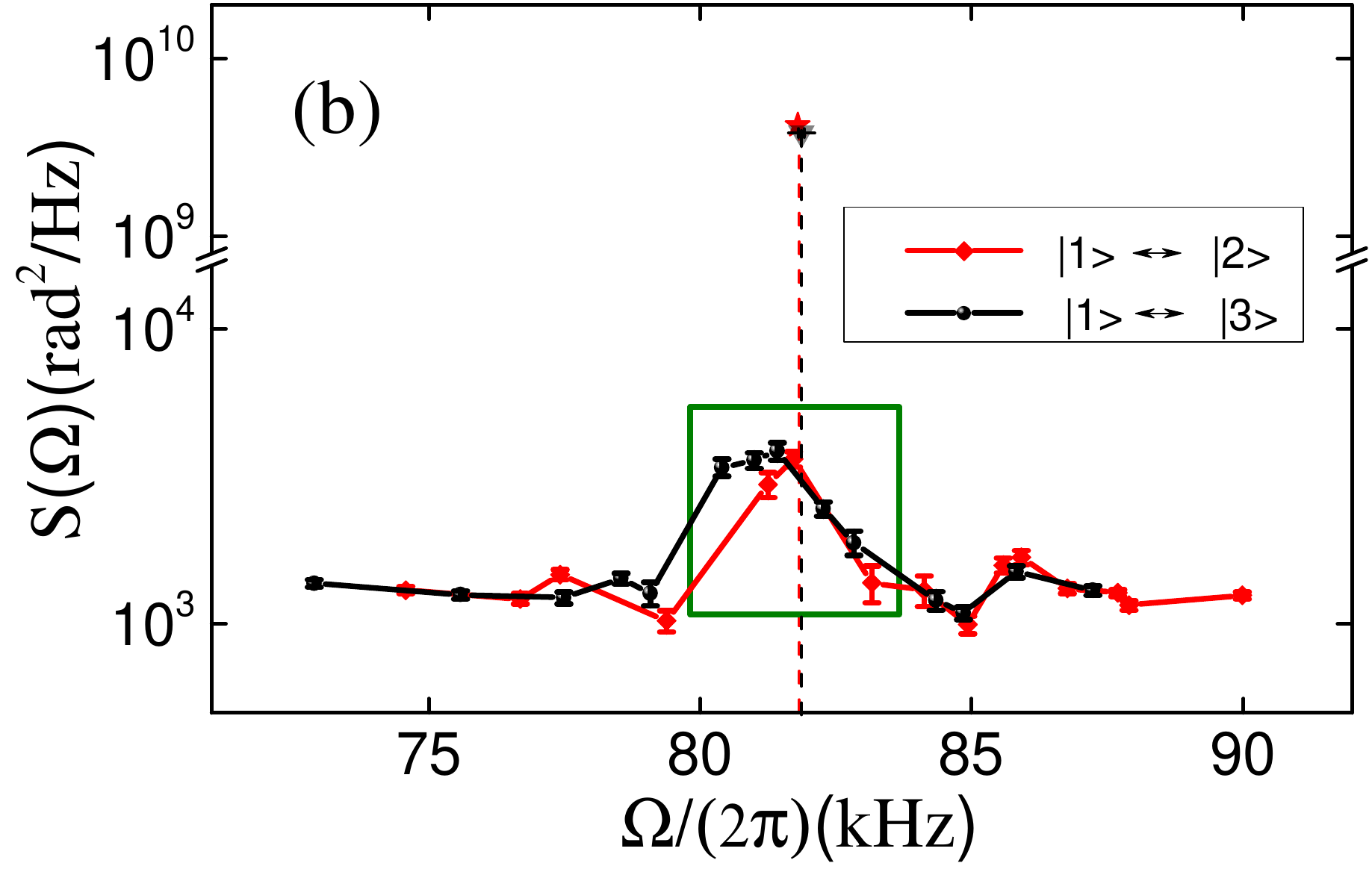}%
\caption{\label{Fig 2}{Decoherence and PSD of dominant noise components. (a) LLN induced Rabi oscillations with different $\Omega$ around $ (2\pi)$82 kHz. The solid curves are fittings using $\widetilde{\Gamma_1}$ and $\widetilde{\Gamma_2}$. Each point represents 200 experiments. (b) Fitting results of $\Omega_{LLN}, \omega_0$ as well as $S(\Omega)$ of the  $(2\pi)$$ 82$ kHz noise component. Each black dot and red diamond represents a fitting result at different $\Omega$. Black triangle and red star  represent the mean values of $\omega_0, \Omega_{LLN}$ for each transition.}}
\end{figure}

The research on a strong coupled environment has already been discussed in some articles like ref.\cite{kotler2013nonlinear} where a discrete spectrum assumption is made. They express the PSD as a sum of discrete noises $S(\omega) = \sum_{k=1}^N S_k \delta (\omega - \omega_k)$ and $S_k$ is the noise strength at $\omega_k$. This method is employed typically when studying some dominant components whose linewidth is relatively small, such as the noise at $ (2\pi)$50 Hz and the harmonics coming from the power line\cite{wang2017single-qubit}. In our system, the $ (2\pi)$82 kHz  and $ (2\pi)$164 kHz components of noise come from the modulation process of the etalon of 729 nm Ti:sapphire laser. Because the electrical signal that drives the etalon has an absolutely narrower linewidth than several hertz, it is reasonable to assume the noise PSD by a $\delta$ function.  However, instead of the PDD technique utilized in ref.\cite{kotler2013nonlinear,wang2017single-qubit} to obtain $S_k$, we propose a new method here to characterize the dominant noise components as well as the weaker coupling region around them where the state evolution is modulated by these components.  We measure the evolution of $P_s$ as a funcion of time $t$ with $\Omega$ being set to around $ (2\pi)$82 kHz, as shown in FIG. \ref{Fig 2}(a). We observe  Rabi oscillations between $\pm X$ states which are similar to the case where a TLS is driven by (an)a (off-)resonant laser beam. Therefore, it is reasonable to make an analogy between  driving lasers and these noise components, which are called laser-like noises (LLN) in the rest of this paper. Generally, the LLN  can be expressed as a form $f_D(t)=E_0\cos(\omega_0t+\phi_0)$ in time domain with the central frequency $\omega_0 \approx (2\pi)82$ kHz or $(2\pi)164$ kHz and amplitude $E_0$ whose exact values need to be extracted from the experimental data. 

\begin{figure}[t]
\centering
\includegraphics[width=0.91\linewidth]{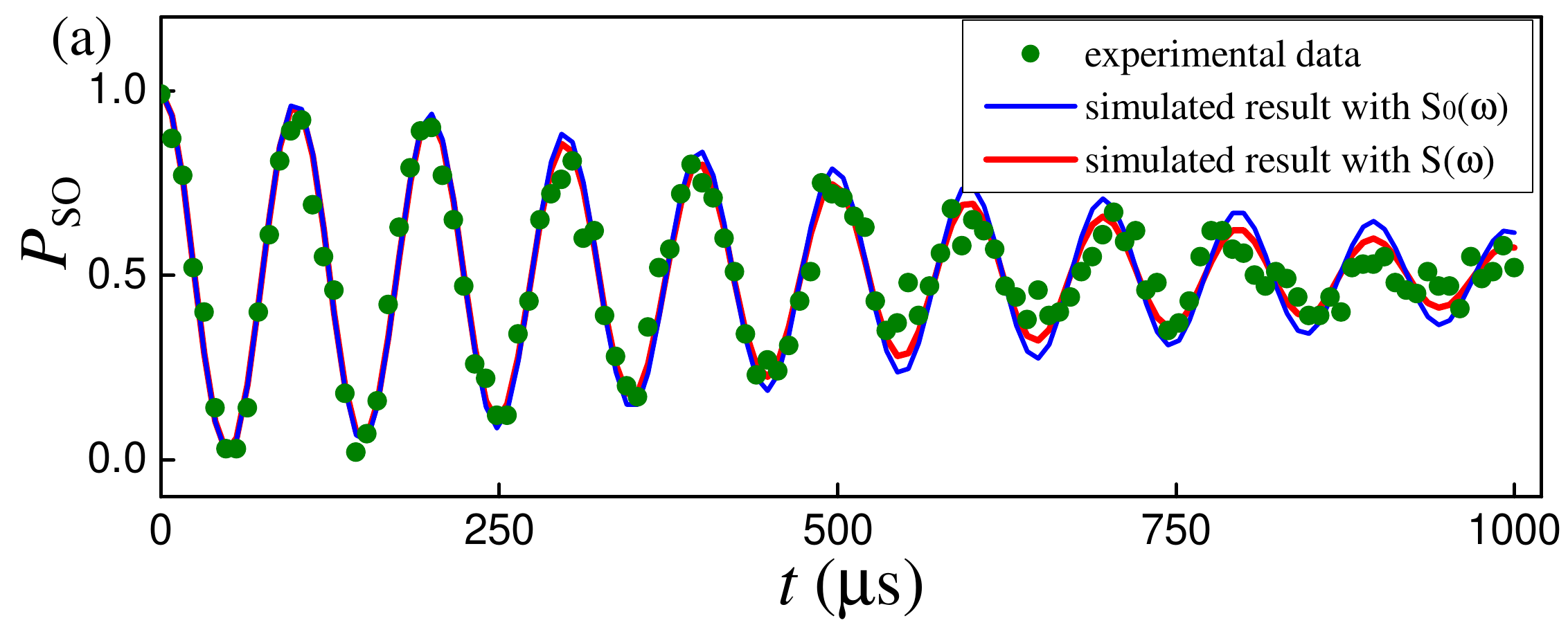}%

\includegraphics[width=0.9\linewidth]{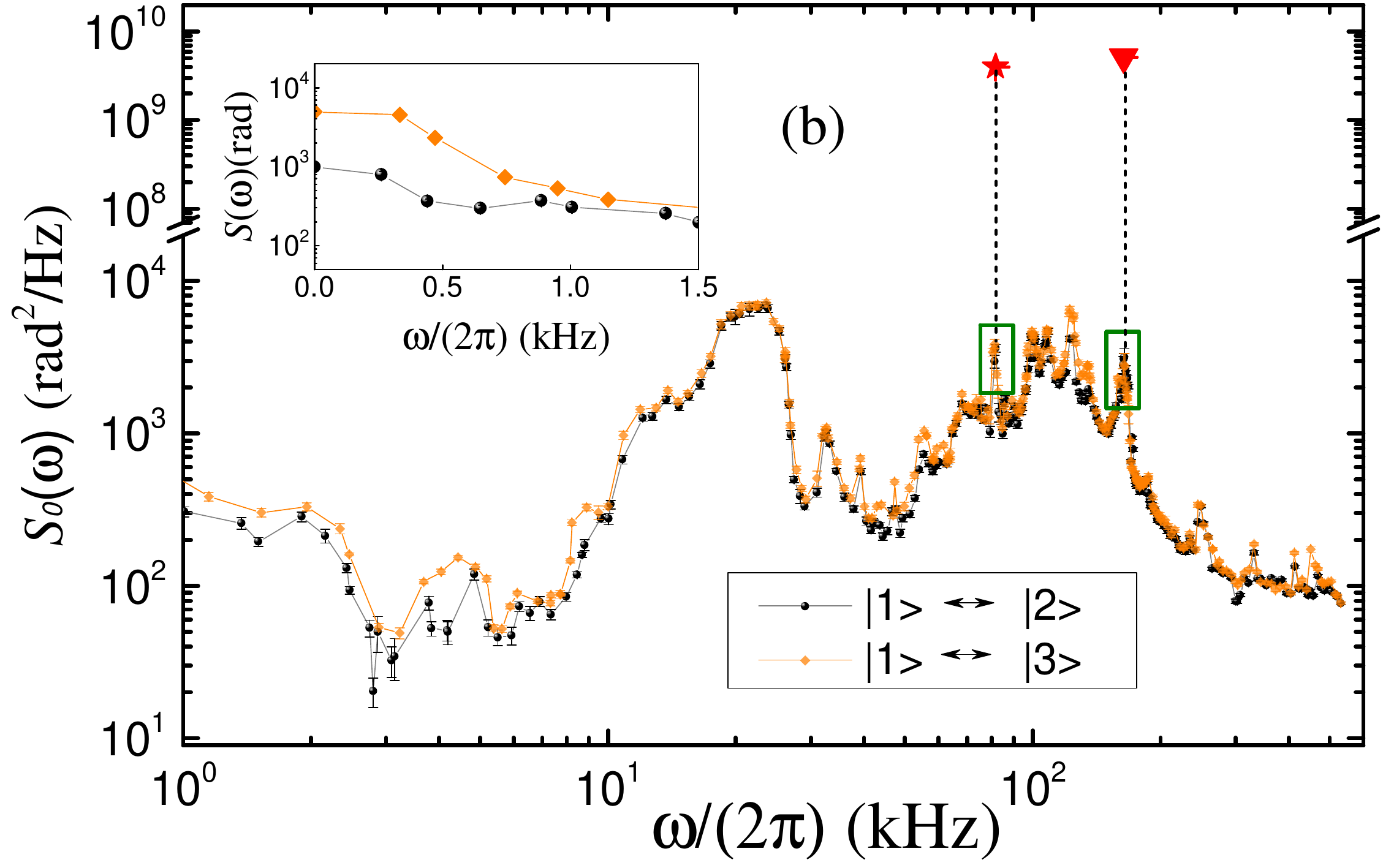}%

\includegraphics[width=0.9\linewidth]{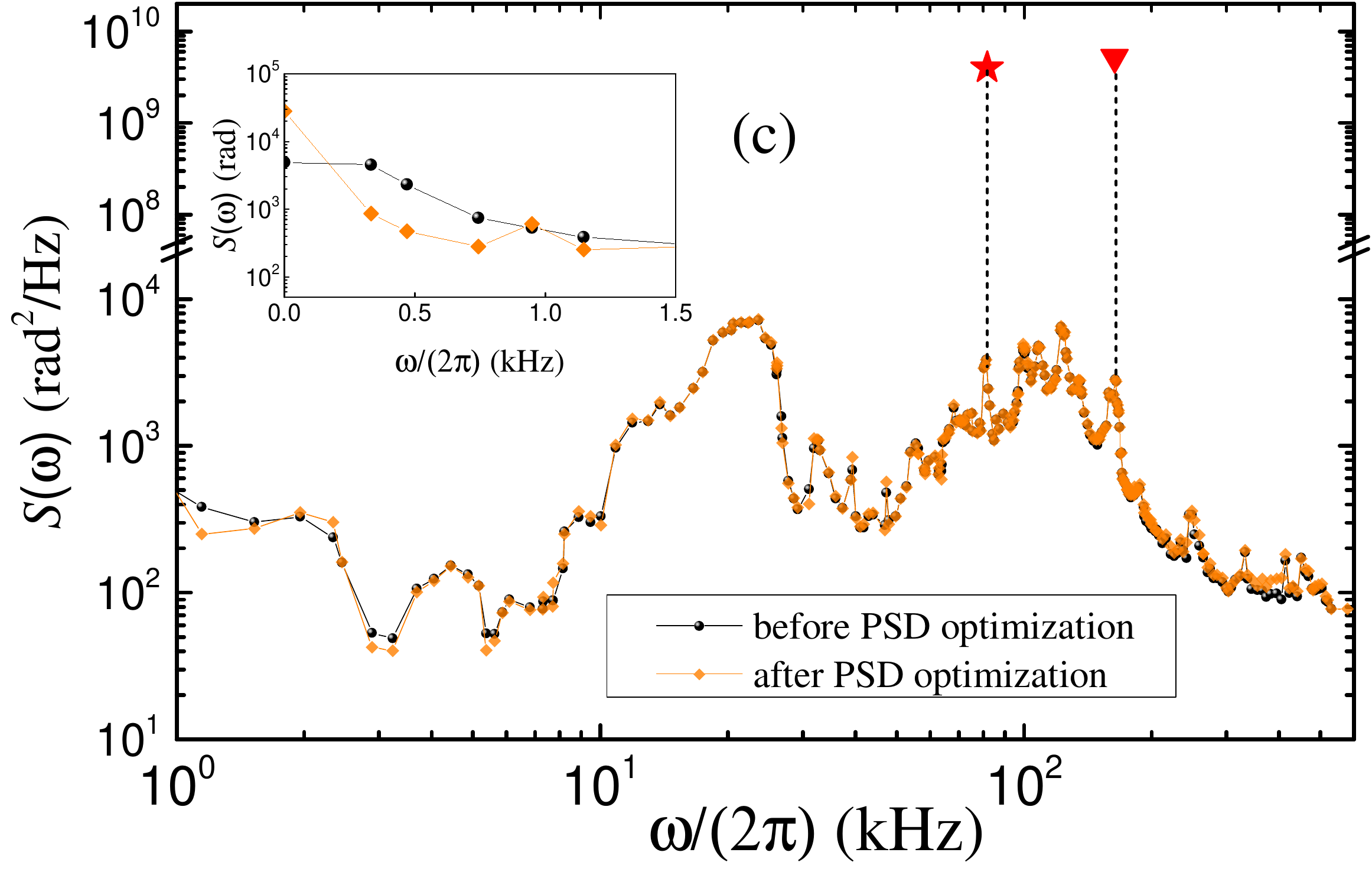}%
\caption{\label{Fig 3}{Noise power spectral density. (a) Survival probability $P_{so}$ of $+X$ state as a function of $t$. The oscillation is resulted from the AC-Stark shift difference explained in the main text. The blue and red curves are numerical simulations using $S_0(\omega)$ and $S(\omega)$ given in (b) and  (c) respectively. Each point represents 200 experiments, and the error bars  are not shown. (b) Experimently determined  noise PSD $S_0(\omega) $ over a range 0-$(2\pi)$530 kHz. The green rectangles enclose the same regions as in FIG. \ref{Fig 2}(b). The inset gives the detailed description for the part below $(2\pi)$1.5 kHz. (c) An example of  $S(\omega)$ extrapolated from $S_0(\omega)$ with the gradient descent protocol for transition $\ket{1}\leftrightarrow \ket{3}$. The inset serves to show remarkable differences between $S_0(\omega)$ and $S(\omega)$ for the frequencies below $(2\pi)$1.5 kHz.}}
\end{figure}

We write the Hamiltonian describing a TLS driven by the LLN in the interaction picture with respect to the frequency of 729 nm laser
\begin{equation}
\label{H^LLN}
H^{LLN} = \frac12 \Omega \sigma_x+ \frac12 \delta \Omega(t) \sigma_x +\frac12 f'(t) \sigma_z+ \Omega_{LLN} \cos(\omega_0 t + \phi_0)\sigma_z.
\end{equation}
Here, $\Omega_{LLN} = E_0/2$ represents the resonant Rabi frequency. $f'(t)$ is the remaining frequency noise that excludes the LLN components from $f(t)$ and $\delta \Omega(t)$ represents the power fluctuations of our 729-nm laser beam. The latter kind of noise is not introduced into the model in Sec.\ref{sectionII} because the exponential decay along $\sigma_x$ axis is immune to it. $\phi_0$ shows the initial phase of LLN of each run and is set as $\phi_0=0$ without introducing errors. We rewrite the static part of $H^{LLN}$ in the rotating frame with $H_0 = \frac12 \omega_0 t\sigma_x$ as:
\begin{equation}
\label{H^LLN_I}
H^{LLN}_{st} = -\frac12 \Delta \sigma_x  + \frac12 \Omega_{LLN}\sigma_z,
\end{equation}
where $\Delta = \omega_0 - \Omega$ is the detuning.  The evolution of a TLS under Eq.(\ref{H^LLN_I}) is a rotation  at the effective Rabi procession frequency $\Omega_R=\sqrt{\Omega_{LLN}^2+ \Delta^2}$. Due to the noise $\frac12 f'(t) \sigma_z$ and  $\frac12 \delta \Omega(t) \sigma_x$, it also experiences a longitudinal decoherence with the relaxation rate $\Gamma_1 = T_1^{-1}=\frac{t}{4\pi}\int_0^\infty  d\omega S(\omega) \rm{sinc}^2[(\omega-\Omega)t/2] \approx S(\Omega)/2$ and pure dephasing with rates $\Gamma_\phi = S_{\delta \Omega}(0)/2$ and $\Gamma_v=S_{\delta \Omega}(\Omega_R)/2$\cite{geva1995on}. $S_{\delta \Omega}(\omega)$ is the PSD of $\delta \Omega(t)$. When $\Delta \not= 0$, the qubit dynamics is conveniently described in a new eigenbasis $\{\pm\widetilde{X}\}$ rotated from the conventional basis $\{ \pm X\}$ by an angle $\eta =\arctan{\frac{\Omega_{LLN}}{\Delta}}$. Here, we define two relaxation rates $\widetilde{\Gamma_1}$ and $\widetilde{\Gamma_2}$ which are analogous to $\Gamma_1$ and $\Gamma_2$ (the transverse relaxation rate). They correspond to the decay of longitudinal and transverse parts of the density matrix in the new basis, respectively.  As a result we obtain\cite{ithier2005decoherence}
\begin{gather}
\label{Gamma1}
\widetilde{\Gamma_1}=  \frac{1+ \rm{cos}^2 \eta}2 \Gamma_1+ \Gamma_v\rm{sin}^2\eta  \\
\label{Gamma2}
\widetilde{\Gamma_2}=\widetilde{\Gamma_1}/2+ \widetilde{\Gamma_\phi} =  \frac{3- \rm{cos}^2 \eta}{4} \Gamma_1 + \Gamma_\phi \rm{cos}^2 \eta + \frac12 \Gamma_v \rm{sin}^2\eta,
\end{gather}
where $\widetilde{\Gamma_\phi} = \frac12 \Gamma_1\rm{sin}^2\eta+ \Gamma_\phi  \rm{cos}^2 \eta$. To simplify the model, we assume $\Gamma_v=0$ because the power of 729-nm laser beam is stabilized in our system. Then we use Eq.(\ref{Gamma1},\ref{Gamma2}) to get the information $S(\Omega)$, $\Omega_{LLN}$ and $\omega_0$ by fitting with the LLN-driving Rabi oscillations, as shown in FIG. \ref{Fig 2}(a). Data are obtained by measuring $P_s$ as a function of $t$. For each  $\Omega$, the LLN results in a Rabi oscillation with angular precession frequency $\Omega_R$. To determine these parameters as precise as possible, we average the outputs of more than twenty Rabi oscillations for each noise component, and finally obtain  $\Omega_{LLN1} =  (2\pi)2.842 \pm (2\pi)0.045$ kHz, $\omega_{01} =  (2\pi)81.832\pm (2\pi)0.055$ kHz for noise around $ (2\pi)82$ kHz and  $\Omega_{LLN2} =  (2\pi)3.362\pm (2\pi)0.081$ kHz, $\omega_{02} =  (2\pi)163.780\pm (2\pi)0.057$ kHz for noise around $ (2\pi)164$ kHz. Therefore, the PSD can be obtained $S(\omega_{01})=\pi \Omega_{LLN1}^2 = (4.007\pm0.001)\times 10^9 \rm{rad}^2$/Hz  and $S(\omega_{02}) = (5.279\pm0.002)\times 10^9 \rm{rad}^2$/Hz. Concerning these results, because the noise at $\omega_{02}$ is a harmonic of the other component, we observe $2\omega_{01}\approx \omega_{02}$. As for $S(\omega_{01})<S(\omega_{02})$, the reason is that the $\omega_{01}$ component is suppressed in our experimental setup. In FIG. \ref{Fig 2}(b), an example of how we get $S(\omega_{01})$ is given. The black triangle and red star show the averaged fitting results  belonging to each transition. It is resonable and convenient to determine the final value just by averaging them.  Besides, the fitting results of $S(\Omega)$ near $\omega_{01}$ are also shown in FIG. \ref{Fig 2}(b). The red diamonds and  black  dots show the noise PSD for transitions $\ket{1}\leftrightarrow \ket{2}$ and $\ket{1}\leftrightarrow \ket{3}$ respectively. Due to the fitting errors, we find that some red diamonds show larger PSD than black dots for the same $\Omega$. In order to get the PSD of both noises, we assume that the PSD of magnetic noise in this region is a constant determined by the values outside the boundary (see Sec. \ref{sectionIIII} and FIG. \ref{Fig 4}).  The green rectangle in FIG. \ref{Fig 2}(b) encloses an area where an increasement appears when $\Omega$ towards $\omega_{01}$. This may be resulted from the power-noise contribution at frequency $\Omega_R$ (i.e. $\Gamma_v$) which is ignored above for simplicity. This means that these data only give the upper limit of $S(\Omega)$. On the other hand, it seems that $\Gamma_v$ has negligible influence on the regions below 80 kHz and above 84 kHz because we observe well matched joint points derived from different fitting approaches in FIG. \ref{Fig 3}(b). This in turn verifies the validity of our model.

\section{Full-spectrum estimation and noise discrimination}
\label{sectionIIII}

On the other hand, we study the rest of the noise spectrum by using the standard continuous dynamical decoupling technique where the evolution of  $P_s$ is described by Eq. (\ref{Ps}). Different from the experiments measuring two dominant noise components above where the 729 nm laser power remains the same in $Y_{\pi/2}$ and driving evolution processes (see the inset in FIG. \ref{Fig 1}(b)), here we use laser pulses with different power in these stages, so that we can save measurement time while the PSD $S(\omega)$ of noise is captured. It can be described in more detiles. We first carefully locate the laser power at a value  where the noise PSD is relatively small based on FIG. \ref{Fig 1}(b), such as $\Omega_S =  (2\pi)200$ kHz whose $\pi/2$ pulse lasts 1.25 $\mu$s. Then this kind of $\pi/2$ pulses are empolyed as $Y_{\pi/2}$ in all experiments where different $\Omega$ are used in the driving evolution processes. In this way, we do not need to calculate and check the $\pi/2$-pulse time when  $\Omega$ is changed. However, it should be mentioned that due to this difference of laser power between $Y_{\pi/2}$ and driving evolution stages, thus different AC-Stark shift, a Ramsey-like oscillation  will be introduced when we monitor $P_s$ as a function of $t$:
\begin{equation}
\label{P_{so}}
P_{so}= \frac12+ \frac12 \cos{(\delta_A t)} \exp(- \int_0^\infty d\omega S(\omega) F(\omega, \Omega, t)),
\end{equation} 
where $\delta_A$ is the AC-Stark shift difference.

FIG. \ref{Fig 3}(a) shows an example of this oscillation decay. The experimental data  are represented by the green dots. We then use the filtering property of the sequence to characterize the noise spectrum. We notice the filter is suffficienty narrow about $\Omega$ so that we can treat the noise as a constant $S(\Omega)$ within its bandwidth $1/t$ and approximate Eq. (\ref{P_{so}}) as $P_{so}= \frac12+ \frac12 \cos{(\delta_A t)} \exp(-\frac{ S(\Omega)}{2}t)$. Utilizing this rectangular approximation approch, the data can be fitted to get $S(\Omega)$. As a result, we obtain the noise PSD $S_0(\omega)$ over a frequency range 0-$(2\pi)$530 kHz by combining other two dominant noise components, as shown in   FIG. \ref{Fig 3}(b). We observe a peak approximately at $ (2\pi)$23.5 kHz belonging to the frequency nosie of laser, and a sharp increase below  $ (2\pi)$1 kHz (the inset) coming from the magnetic field fluctuations. We also observe a stronger noise for transition $\ket{1}\leftrightarrow \ket{3}$ (orange diamonds) than $\ket{1}\leftrightarrow \ket{2}$ (black dots), which is consistent with our assumption. 

To extend the accuracy of the measured noise PSD, we propose a gradient descent protocol based on matching to the experimental oscillation decay curves. Define a objective function as the sum of the squared error between the experimentally measured decay $P_s(t_j)$ and the calculated decay rule $P_s'(t_j) = \frac12+ \frac12 \cos{(\delta_A t)} \exp(- \int_0^\infty d\omega S'(\omega) F(\omega, \Omega, t_j))$ for a given $S'(\omega)$:
\begin{equation}
\label{J}
J  =  \sum_{j=1}^M( P_s(t_j)-  P_s'(t_j))^2.
\end{equation} 
We then calculate the gradient of this function $\frac{\partial J}{\partial S'(\omega_i)}$ for any target frequency $\omega_i$. The gradient is used to update the initial $S'(\omega)$ towards a closer matching of all the experimental and calculated decays. Finally, a comparison between the optimized PSD  $S(\omega)$ (orange diamonds) and $S_0(\omega)$ (black dots) corresponding to  $\ket{1}\leftrightarrow \ket{3}$ is shown in FIG. \ref{Fig 3}(c). We can easily find that the gradient optimization has the most remarkable effects for the extreme points of $S(\omega)$, especially the ones between sharp increases and decreases. The reason is simple. In these intervals, the rectangular approximation approch may not be valid anymore because $S(\omega)$ cannot be replaced by $S(\Omega)$ within the filter bandwidth. After obtaining the optimized $S(\omega)$, we also verify its correctness. In FIG. \ref{Fig 3}(a), we give the simulated evolution $P_{so}$ based on $S_0(\omega)$ (blue curve) and $S(\omega)$ (red curve) respectively. It is obvious that the red cuve matches the data points better, which shows a greater fidelity of $S(\omega)$.

\begin{figure}[h]
\centering
\includegraphics[width=0.9\linewidth]{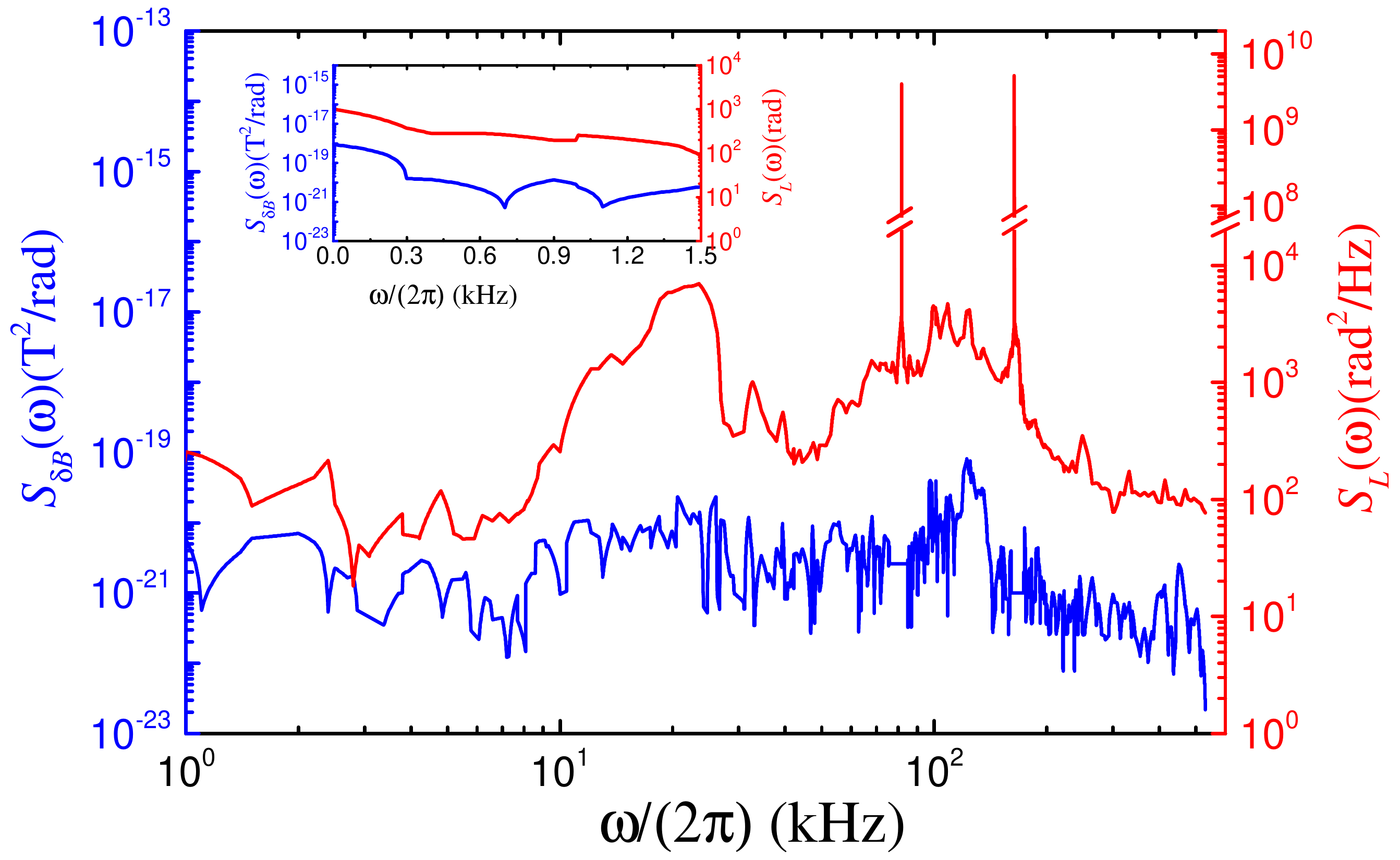}%
\caption{\label{Fig 4}{Noise discrimination. The PSD of laser and magnetic field noises are represented by the red and blue lines respectively. The inset shows the details of PSD below $(2\pi)$1.5 kHz.}}
\end{figure}

To get the PSD of laser frequency noise, we utilize the $S(\omega)$ for both transitions  $\ket{1}\leftrightarrow \ket{2}$ and  $\ket{1}\leftrightarrow \ket{3}$ obtained above whose difference can be used to speculate the magnetic field fluctuations. According to the atomic physics, the magnetic field noise induced energy level shift changes the transition circular frequency by $f_B(t) = \frac{\mu_B \delta B(t)}{\hbar}[g_j(D_{5/2})m'-g_j(S_{1/2})m]$, where $\mu_B$ is the Bohr magneton, $g_j$  the Lande g factors, and $m'$, $m$ are the magnetic quantum numbers of the $D_{5/2}$ and $S_{1/2}$ state, respectively. $\delta B(t)$ represents the magnetic field strength fluctuations. For ${}^{40}Ca^+$, $g_j(S_{1/2})  \approx 2$ and $g_j(D_{5/2}) \approx 1.2$, so that the PSD of $\delta B(t)$ can be expressed as  $S_{\delta B}(\omega) = [S_{1\leftrightarrow3}(\omega)-S_{1\leftrightarrow2}(\omega)]\hbar^2/(24\times 0.16 \mu_B^2)$. And next we obtain the PSD of laser frequency noise by $S_L(\omega)  = S_{1\leftrightarrow2}(\omega) - [S_{1\leftrightarrow3}(\omega)-S_{1\leftrightarrow2}(\omega)]/24 $, as shown in FIG. \ref{Fig 4}. From the discrete data points in FIG. \ref{Fig 3}(c) to the continuous lines here, we have employed the interpolation method.  In FIG. \ref{Fig 4}, we observe complex PSD for both kinds of noise which cannot be described by a single linetype, such as $1/\omega^\alpha$ used in many articles. We also see a small  interconnected trend between $S_{\delta B}(\omega)$ and $S_L(\omega)$ in some peaks, such as $\omega \approx (2\pi)23.5$ kHz. In our opinions, this is primarily derived from the limited  accuracy at these points where the fitting error is considerable. 

\section{Compared to the beat-note scheme}
\label{sectionIIIII}
\begin{figure}[h]
\centering
\includegraphics[width=0.9\linewidth]{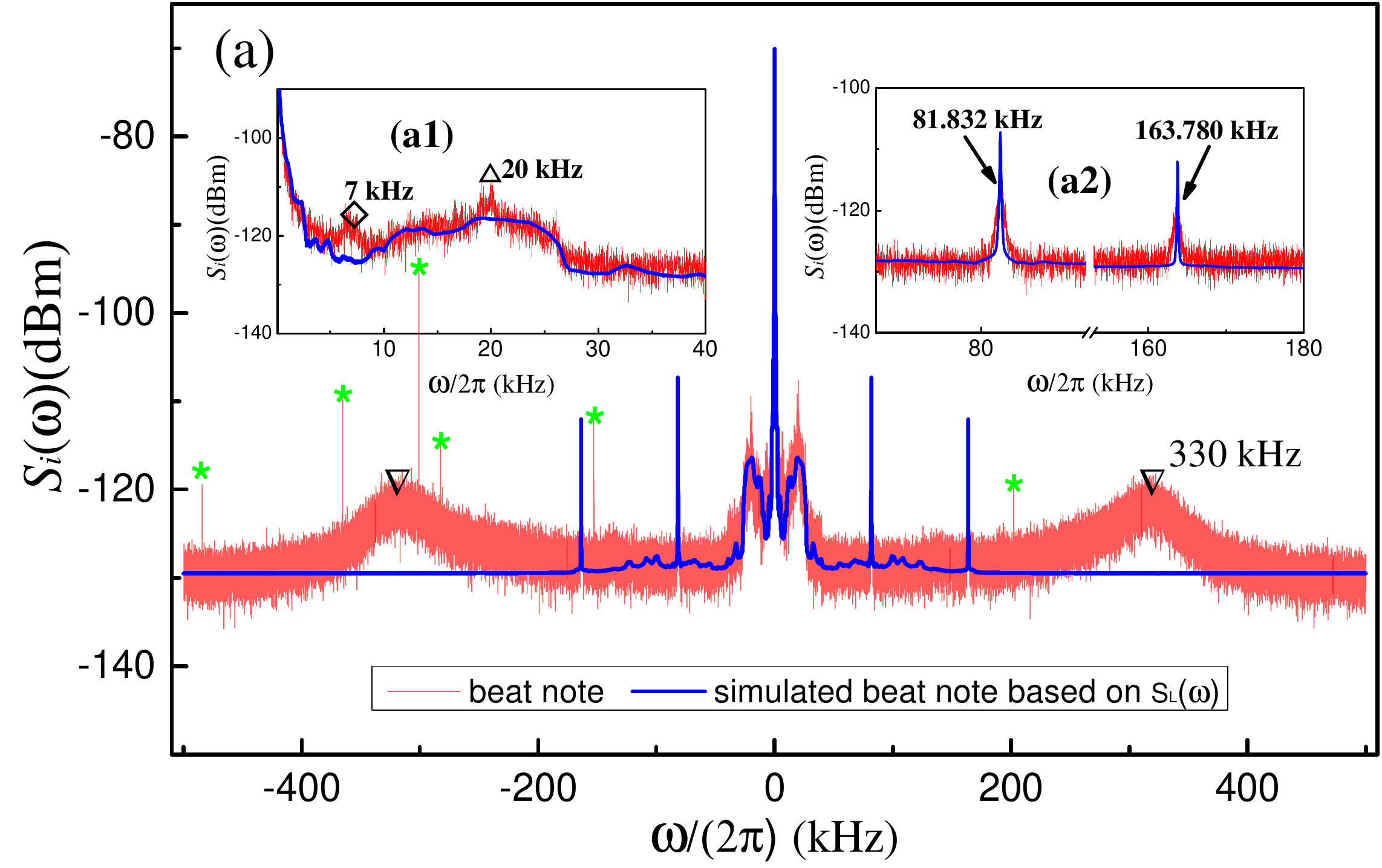}%

\includegraphics[width=0.89\linewidth]{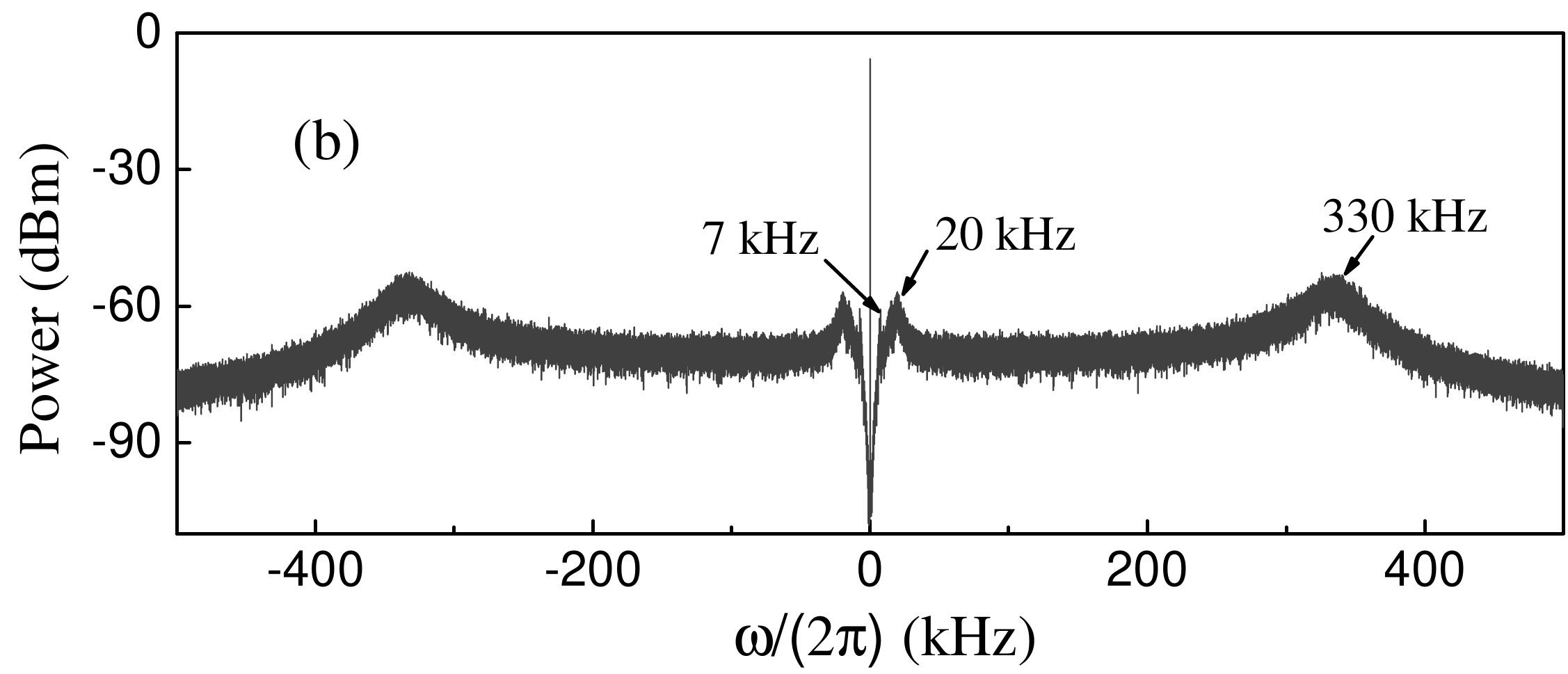}%
\caption{\label{Fig 5}{(a) Electrical spectra of the (simulated) laser beat-note signals. Red curves show the beat note between our Ti:sapphire laser under test and a diode laser at 729 nm, while the blue curves represent the simulated beat-note result according to Eq. (\ref{Si}) by setting $S_1(\omega)=S_L(\omega)$ and $S_2(\omega)=0$. Note that there are several peaks marked with green stars. They are from the photodiode instead of two laser beams. Usually, the beat note is determined by the noise from both lasers. This is exactly the reason why two curves  have a divergent behaviour at $\pm$7 kHz (diamond in inset (a1)), $\pm$20 kHz (triangle in inset (a1)) and $\pm$330 kHz (upside-down triangles) where the frequency noise from diode laser dominates. Insets give more details for the region (a1) 0.1 kHz $\sim$ 40 kHz and (a2) 70 kHz $\sim$ 180 kHz. We see a good matching between two kinds of curves, such as the dominant noise components centered at 81.832 kHz and 163.5 kHz in inset (a2). (b) The PDH error spectrum of the 729-nm diode laser measured by a rf spectral analyzer.}}
\end{figure}

Alternatively, in order to characterize the frequency noise of this Ti:sapphire laser, we beat it with a distinct diode laser at 729 nm. The  electrical spectrum of the beat-note signal is presented in FIG.   \ref{Fig 5}(a), as the red curves. In fact, they are PSD of the output electrical signal $i$ of the photodiode which is used to convert the combined bichromatic light field $($the amplitude of the target(reference) laser is $E_1(E_2))$ to the photocurrent. The carrier frequency of the beat note has been shifted to 0 Hz from 5.36 MHz which represents the frequency difference of two lasers. Note that due to the dominant noise  components from photodiode, there are several asymmetrical spikes with respect to 0 Hz. And to make it clearer, we mark them with green stars  in the figure.

In fact, we can find the relation between the beat-note signal $S_i(\omega)$ and the frequency noise PSD of lasers by the equation\cite{stephan2005laser}
\begin{equation}
\begin{split}
\label{Si}
S_i(\omega)=(\frac{\alpha E_1 E_2}{2})^2 \int_{-\infty}^\infty d \tau \cos{(\omega-\omega_0)\tau} \\
\cdot \exp{-\frac{\tau^2}{4 \pi}\int_{-\infty}^\infty d \omega' (S_1(\omega')+S_2(\omega'))\rm{sinc}^2(\omega' \tau/2)},
\end{split}
\end{equation}
where  $\omega_0$ is the frequency difference of two lasers. $\alpha$ shows the photoelectric conversion coefficient of the photodiode. $S_1(\omega)$ and $S_2(\omega)$ represent the noise PSD of laser under test and reference laser  respectively. Eq. (\ref{Si}) implies that the beat-note result in FIG. \ref{Fig 5}(a) is determined by the noise of both lasers. And from this beat note, we can not tell the contributions of each laser apart if we do not have any prior information, just as what  we said in section \ref{sectionI}. To make a comparison, let us onsider a case where the target laser is beat with an ideal laser, i.e. $S_1(\omega)=S_L(\omega)$ and $S_2(\omega)=0$, then we can obtain a simulated beat-note result, as the blue curves shown in FIG. \ref{Fig 5}(a). This simulated beat note actually only contains the frequency noise information of the target Ti:sapphire laser. By comparing it with the experimental beat note in FIG. \ref{Fig 5}(a), we observe that the two curves match very well except for three distinct components centered at $\pm$7 kHz (marked as a diamand in inset (a1)), $\pm$20 kHz (the triangle in inset (a1)) $\pm$330 kHz (upside-down triangles in the main plot). Because these components have been proved to be from the diode laser according to the PDH error signal spectrum (FIG. \ref{Fig 5}(b)) measured by a rf spectrum analyzer (for more information about the PDH error signal spectrum, see ref. \cite{text}), we conclude that the validity of our model and result expressed in section \ref{sectionIII}$\sim$\ref{sectionIIII} are confirmed. In order to show more details in low frequency regions, we also provide two insets showing a good matching between two curves in the range (a1) 0.1 kHz $\sim$ 40 kHz and (a2) 70 kHz $\sim$ 180 kHz. In inset (a2) we find that the central frequencies of  dominant noise components at 81.832 kHz and 163.780 kHz  are  accurately estimated in section \ref{sectionIII}, while the strength seems to be less convincing because the peaks of the beat note are wider and lower. In fact, this property is resulted from the convolution with other noise components such as the one at 330 kHz from the diode laser.  Therefore the validity of our result should not be denied just by this differencce. And to be more precisely, this actually exposes a shortcoming of the beat-note method.

\section{Conclusion}
\label{sectionIIIIII}
We have employed the continuous dynamical decoupling protocol in this work to estimate the laser frequency noise spectrum. It is based on controlling the system-enviroment interaction by applying a continuous driving field along $\sigma_x$ axis to the system after initializing the system on $+X$ state. This driving field and the corresponding Rabi frequency $\Omega$ create the dressed states $+X$ and $-X$ between which the transition frequency is $\Omega$ in the interaction picture. Because the noise $f(t)$ works along the $\sigma_z$ axis, it actually causes a state transition between $+X$ and $-X$ states. In this way, the system will suffer less from the most noise parts of frequency bands except for those around $\Omega$  as long as the noise spectrum $S(\Omega)$ is nonnegligible.  By using this filtering property of controlling sequences and the gradient descent algorithm, we have directly characterized the overall noise PSD up to $(2\pi)$530 kHz, which in turn enables the design of coherent control that targets a specific noise component. However, in our case of laser noise, there are two dominant components centered at $(2\pi)$81.832 kHz and $(2\pi)$163.780 kHz that are too strong to use this filter function method. Instead, we propose a new approach which regards these noises as driving lasers performed in the space spanned by the dressed states $\pm X$  after being initialized on $+X$. To simplify the model, a discreteness assumption is made. The LLN induced decay Rabi oscillation in the $\{+X, -X\}$ space  can give information about the central frequency and strength of this noise. On the other hand, due to the equivalent timescales of  both laser noise and magnetic field noise acting on system, we finally obtain PSD of each by encoding the qubit on different Zeeman sublevels of ${}^2D_{5/2}$. Especially,  this method is of great  significance for characterizing the laser noise from the mixed noise. And by comparing with the beat-note signal, our result is verified.  This indicates that we do not need to employ the optical heterodyne approach anymore for which at least hundreds of kilometers of optical fibers or several similar lasers are required.

\begin{acknowledgments}
This work is supported by  the National Basic Research Program of China under Grant
No. 2016YFA0301903 and the National Natural Science Foundation of China under Grant No. 61632021, No.11904402 and No.12004430.
\end{acknowledgments}

\bibliographystyle{apsrev4-1}

\begin{thebibliography}{41}
\bibitem [{tex()}]{text}%
  \BibitemOpen
  \href@noop {} {}\bibinfo {note} {The error signal in PDH frequency
  stabilization process represents the nonlinear response to the laser
  frequency fluctuations and can be analyzed by a rf spectral analyzer.
  However, this noise analysis method is usually not employed as a direct and
  precise measurement approach of noise PSD for two reasons. First, its
  spectrum is different from the frequency noise PSD, and only gives general
  information about the noise components, such as the position and modulated
  relative strength. Second, due to the limitations of rf spectral analyzer,
  this method cannot provide a precise measurement of noise in low frequency
  region. For example, the operating range of our rf spectral analyzer
  (KEYSIGHT: CXA Signal Analyzer N9000B) is 9 kHz to 3 GHz, so the noise
  components below 9 kHz cannot be measured with gratifying
  accuracy.}\BibitemShut {Stop}%
\end{thebibliography}
\providecommand{\noopsort}[1]{}\providecommand{\singleletter}[1]{#1}%
%




\end{document}